\documentclass[reprint,nofootinbib, superscriptaddress]{revtex4-1} 

\usepackage{graphicx}
\usepackage{color}
\usepackage{amsmath}
\usepackage{amssymb}
\usepackage{bm}
\usepackage{dcolumn}
\usepackage{placeins}
\usepackage[pdfusetitle]{hyperref}
\usepackage[anythingbreaks]{breakurl}
\definecolor{link}{rgb}{0.07, 0.07, 0.80}
\hypersetup{
  colorlinks=true,
  linkcolor=link,
  citecolor=link,
  urlcolor=link
}
\usepackage{caption, subcaption}
\usepackage{sidecap}
\sidecaptionvpos{figure}{c}

\usepackage{txfonts}

\makeatletter
\let\intertext=\shortintertext
\everydisplay{\medmuskip=1.2mu\relax}
\makeatother

\begin{document}

\preprint{APS/123-QED}

\title{Probing photoisomerization processes by means of multi-dimensional electronic spectroscopy:\\
  The multi-state quantum hierarchical Fokker-Planck equation approach}

\author{Tatsushi Ikeda}
\email{ikeda.tatsushi.37u@kyoto-u.jp}
\affiliation{Department of Chemistry, Graduate School of Science, Kyoto University, Kyoto 606-8502, Japan}
\author{Yoshitaka Tanimura}
\email{tanimura.yoshitaka.5w@kyoto-u.jp}
\affiliation{Department of Chemistry, Graduate School of Science, Kyoto University, Kyoto 606-8502, Japan}

\date{\today}

\begin{abstract}
Photoisomerization in a system with multiple electronic states and anharmonic potential surfaces in a dissipative environment is investigated using a rigorous numerical method employing quantum hierarchical Fokker-Planck equations (QHFPE) for multi-state systems.
We have developed a computer code incorporating QHFPE for general-purpose computing on graphics processing units (GPGPU) that can treat multi-state systems in phase space with any strength of diabatic coupling of electronic states under non-perturbative and non-Markovian system-bath interactions.
This approach facilitates the calculation of both linear and nonlinear spectra.
We computed Wigner distributions for excited, ground, and coherent states.
We then investigated excited state dynamics involving transitions among these states by analyzing linear absorption and transient absorption processes and multi-dimensional electronic spectra with various values of the heat bath parameters.
Our results provide predictions for spectroscopic measurements of photoisomerization dynamics.
The motion of excitation and ground state wavepackets and their coherence involved in the photoisomerization were observed as the profiles of positive and negative peaks of two-dimensional spectra.
\end{abstract}

\maketitle

\section{Introduction}
\label{sec:introduction}
Analysis of photoinduced chemical processes has played important roles in the study of reaction dynamics in condensed phases.
Examples of such processes that have been investigated with spectroscopic techniques include photoisomerization \cite{waldeck1991cr,ishii2004cpl,takeuchi2008science,briney2010jpca,hujino2001jpca,schultz2003jacs,saito2003om,kandori2001bc,kobayashi2001nature,ruhman2002jacs,hellingwerf2003jpca,kuramochi2012jpcl,conyard2017nc,conyard2014jacs}, photodissociation \cite{zewail1994book}, molecular photoswitching \cite{feringa2011book,mammana2011jpcb}, and exciton and charge-transfer dynamics \cite{may2008book}.
Systems exhibiting such processes have been described using models consisting of several electronic states strongly coupled to primary nuclear coordinates \cite{mukamel1999book}.
In such models, environmental effects that arise from solvation, interaction with protein molecules, etc.
are generally described by an interaction between the primary nuclear coordinates and a harmonic oscillator bath \cite{tanimura1994jcp,tanimura1997jcp,maruyama1998cpl,tanimura2006jpsj,ishizaki2009jcp,cho2007acp,fuller2014nc}.
Typically, the potential of the primary coordinates is assumed to be harmonic.
This allows the number of nuclear coordinate degrees of freedom to be reduced using a cumulant expansion method \cite{mukamel1999book} or a path integral method \cite{garg1985jcp,tanimura1993pre}.
While the harmonic approximation applied to the nuclear coordinates works well for the description of intramolecular and intermolecular motion, a more rigorous approach is needed to describe reaction dynamics such as photoisomerization and photodissociation, in which there is a change in the configuration of the molecule that takes places along the molecular coordinates.
In this paper, we investigate nonlinear optical responses in photoisomerization processes by explicitly taking into account the anharmonic forms of actual nuclear potentials.

Experimental investigations of excited state dynamics in photoisomerization employing lasers have been carried out for stilbene \cite{waldeck1991cr,ishii2004cpl,takeuchi2008science,briney2010jpca} and azobenzene \cite{hujino2001jpca,schultz2003jacs,saito2003om} in various solvents, retinal in bacteriorhodpsin \cite{kandori2001bc,kobayashi2001nature,ruhman2002jacs}, and a photoactive yellow protein (PYP) system \cite{hellingwerf2003jpca,kuramochi2012jpcl}.
Due to the complexity of such systems, experimental investigations of isomerization processes that they exhibit require significant theoretical input with regard to the time evolution of the system under laser excitation.
Theoretically, the excited state dynamics of such systems that take into account nuclear degrees of freedom and electronic states, have been investigated approaches using equation of motion \cite{zusman1980cp,shi2009jcp,sheu1998jcp,donoso2000jcp,kuhn2000jcp,thoss2000jcp,feskov2005jcp,egorova2008jcp,kowalewski2017sd}, approaches that employ surface hopping \cite{tully1990jcp,petit2014jcp}, mixed quantum-classical dynamics methods \cite{kapral1999jcp,puzari2004jcp,hanna2008jcp,rank2010jcp,hsieh2012jcp,raymond2015jpcm,ando2003jcp,xie2016jpca}, and ab initio multiple spawning methods \cite{levine2007arpc}.
These approaches, however, do not have the capability of properly treating electronic coherence.
Electronic coherence plays an essential role in the study of nonlinear optical response in photoisomerization processes.
Here, we study photoisomerization processes using multi-state hierarchical quantum Fokker-Planck equations of motion (MS-QHFPE) approach \cite{tanimura2006jpsj,tanimura1997jcp,maruyama1998cpl}, which is an extension of reduced hierarchical equations of motion for open quantum dynamics \cite{tanimura1989jpsj,tanimura1990pra,tanimura1991pra,tanimura1992jcp,ishizaki2005jpsj, tanimura2014jcp,tanimura2015jcp,chen2010jcp,chen2011jcp,hein2012njp,kreisbeck2013jpcb,tanimura1994jcp,tanimura2012jcp,gelin2013jcp,dijkstra2015jcp,tanaka2009jpsj,tanaka2010jcp,ding2012jcp,liu2014jcp}.
This approach allows us to study the dynamics of quantum open systems as well as nonlinear optical spectra in a numerically rigorous manner.
While this technique can be used to treat systems with arbitrary potentials, due to the existence of nonlocal potential terms in Wigner space, solving the kinetic equations is extremely numerically demanding.
For this reason, we employ an efficient algorithm to evaluate the quantum Liouville term and adapt general-purpose computing on graphics processing units (GPGPU) in order to compute the nonlocal potential terms efficiently.

This paper is organized as follows.
In Sec.~\ref{sec:model}, we introduce a model of photoisomerization in a condensed phase.
We then illustrate the MS-QHFPE.
In Sec.~\ref{sec:optical_observables}, we present results for linear and nonlinear optical observables calculated using the model introduced in Sec.~\ref{sec:model} and analyze these profiles.
Section~\ref{sec:conclusion} is devoted to concluding remarks.

\section{The multi-state quantum hierarchical Fokker-Planck equations}
\label{sec:model}
We consider a molecular system with multiple electronic states, $|j\rangle $, strongly coupled to a single primary reaction coordinate.
The Hamiltonian of the system is given by \cite{mukamel1999book,tanimura1994jcp,tanimura1997jcp,maruyama1998cpl,tanimura2006jpsj,ishizaki2009jcp,cho2007acp}
\begin{align}
  \hat{\bm{H}}_{\mathrm{S}}&=\frac{\hat{p}^{2}}{2m}\otimes \bm{I}+\sum _{j,k}|j\rangle U_{jk}(\hat{q})\langle k|,
  \label{eq:system-Hamiltonian}
\end{align}
where $\bm{I}\equiv \sum _{j}|j\rangle \langle j|$, and $\hat{q}$, $\hat{p}$, and $m$ are the primary reaction coordinate, conjugate momentum, and mass, respectively.
The diagonal element $U_{jj}(q)$ represents the diabatic potential surface of the $j$th electronic state.
The off-diagonal element $U_{jk}(q)$ represents the diabatic coupling between the $j$th and $k$th states.
For the multi-state system described by Eq.~\eqref{eq:system-Hamiltonian}, the density matrix can be expanded in the electronic basis set as
\begin{align}
  \hat{\bm{\rho }}(q,{q}';t)&=\sum _{j,k}|j\rangle \rho _{jk}(q,{q}';t)\langle k|,
  \label{eq:system-densitymatrix}
\end{align}
where $\rho _{jk}(q,{q}';t)$ is the density matrix for the electronic density states $j$ and $k$ expressed in the coordinate representation.

The molecular system is also coupled to a bath, which is represented by a set of harmonic oscillators, with the $n$th bath oscillator possessing frequency $\omega _{n}$, mass $m_{n}$, coordinate $\hat{x}_{n}$, and momentum $\hat{p}_{n}$.
The interaction between the primary nuclear coordinate and the $n$th bath oscillator is assumed to be linear concerning $\hat{x}_{n}$, with a coupling strength $g_{n}$.
The total Hamiltonian is then given by
\begin{align}
  \hat{\bm{H}}_{\mathrm{tot}}&=\hat{\bm{H}}_{\mathrm{S}}+\hat{\bm{H}}_{\mathrm{B+I}},
  \label{eq:total-Hamiltonian}
\end{align}
where
\begin{align}
  \hat{\bm{H}}_{\mathrm{B+I}}&=\sum _{n}\left\{\frac{\hat{p_{n}}^{2}}{2m_{n}}+\frac{m_{n}{\omega _{n}}^{2}}{2}\left(\hat{x}_{n}-\frac{g_{n}V(\hat{q})}{m_{n}{\omega _{n}}^{2}}\right)^{2}\right\}\otimes \bm{I},
  \label{eq:bath-Hamiltonian}
\end{align}
and $V(q)$ represents the system side of the system-bath coupling function.
Although, in this paper, we restrict our analysis to the case of linear interaction, $V(q)=q$, here we treat $V(q)$ as an arbitrary function for future expansion to include the effects of anharmonic mode-mode coupling and vibrational dephasing \cite{tanimura2000jpsj,ishizaki2006jcp,sakurai2011jpca,ikeda2015jcp,ito2016jcp}.
The noise arising from the bath is characterized by cumulants expressed in terms of the multi-body correlation functions of the collective bath coordinate, $\hat{X}\equiv \sum _{n}g_{n}\hat{x}_{n}$.
Because the bath consists of harmonic oscillators and the system-bath interaction is linear, the higher-order cumulants vanish, and the influence of the bath is solely determined by the symmetrized correlation function $C_{\mathrm{B}}(t)\equiv \langle \hat{X}(t)\hat{X}(0)+\hat{X}(0)\hat{X}(t)\rangle _{\mathrm{B}}/2$, where $\langle \dots \rangle _{\mathrm{B}}$ represents the thermal average over the bath modes.
This function satisfies the fluctuation-dissipation relation $C_{\mathrm{B}}(\omega )=\hbar \omega /2\coth \left(\beta \hbar \omega /2\right)\Psi _{\mathrm{B}}(\omega )$, for the relaxation function $\Psi _{\mathrm{B}}(t)\equiv \int _{t}^{\infty }\mathrm{d}t'\Phi _{\mathrm{B}}(t')$, where $\Phi _{\mathrm{B}}(t)\equiv (i/\hbar )\langle [\hat{X}(t),\hat{X}(0)]\rangle _{\mathrm{B}}$ is the response function.
In general, the fluctuation-dissipation theorem is an approximate relation that is valid only in the case of a weak system-bath interaction.
However, for the present harmonic bath with a linear system-bath interaction, this relation holds for any system-bath coupling strength.

The bath is characterized by the spectral distribution function, defined by
\begin{align}
  J(\omega )&\equiv \sum _{n}\frac{{g_{n}}^{2}}{2m_{n}\omega _{n}}\delta (\omega -\omega _{n}),
\end{align}
and the inverse temperature, $\beta \equiv 1/k_{\mathrm{B}}T$, where $k_{\mathrm{B}}$ is the Boltzmann constant.
The relaxation function and the symmetrized correlation function can be expressed in terms of $J(\omega )$ and $\beta $ as
\begin{align}
  \Psi _{\mathrm{B}}(t)&=2\int _{0}^{\infty }\mathrm{d}\omega \frac{J(\omega )}{\omega }\cos \omega t
  \label{eq:bath-response}\\
  \intertext{and}
  C_{\mathrm{B}}(t)&=\hbar \int _{0}^{\infty }\mathrm{d}\omega J(\omega )\coth \left(\frac{\beta \hbar \omega }{2}\right)\cos \omega t.
  \label{eq:bath-correlation}
\end{align}
When each function given in Eqs.~\eqref{eq:bath-response} and \eqref{eq:bath-correlation} is expressed as a linear combination of exponential functions and the Dirac's delta function as
\begin{align}
  \Psi _{\mathrm{B}}(t)&=\sum _{k=0}^{K}r_{k}\cdot \gamma _{k}e^{-\gamma _{k}\left|t\right|}
  \label{eq:expansion-of-bath-relaxation}
  \intertext{and}
  C_{\mathrm{B}}(t)&=\sum _{k=0}^{K}c_{k}\cdot \gamma _{k}e^{-\gamma _{k}\left|t\right|}+c_{\delta }\cdot 2\delta (t),
  \label{eq:expansion-of-bath-correlation}
\end{align}
which is realized in the Drude form for exponentially decaying noises \cite{tanimura2006jpsj,tanimura1997jcp,maruyama1998cpl,tanimura1989jpsj,tanimura1990pra,tanimura1991pra,tanimura1992jcp,ishizaki2005jpsj,tanimura2014jcp,tanimura2015jcp, chen2010jcp,chen2011jcp,hein2012njp}, Lorentz form for protein environments \cite{kreisbeck2013jpcb}, and Brownian form for electronic excitation and electron transfer cases \cite{tanimura1994jcp,tanimura2012jcp,gelin2013jcp,dijkstra2015jcp,tanaka2009jpsj,tanaka2010jcp,ding2012jcp,liu2014jcp}, we can obtain reduced equations of motion for the density matrix of the system in the form of the hierarchical equations of motion (HEOM).
Here, we assume that $J(\omega )$ possesses the Drude form,
\begin{align}
  J(\omega )=\frac{m\zeta }{\pi }\frac{\omega \gamma ^{2}}{\gamma ^{2}+\omega ^{2}},
  \label{eq:Drude-J}
\end{align}
where $\zeta $ is the system-bath coupling strength, and $\gamma $ represents the
width of the spectral distribution.

We now introduce the Wigner distribution function, which is the quantum analogy of the classical distribution function in phase space.
Although computationally expensive, the HEOM in Wigner space is ideal for studying quantum transport systems, because it allows for the treatment of continuous systems, utilizing open boundary conditions and periodic boundary conditions \cite{tanimura1992jcp}.
In addition, the formalism can accommodate the inclusion of an arbitrary time-dependent external field \cite{tanimura1997jcp,maruyama1998cpl}.
Because we can compare quantum results with classical results obtained in the classical limit of the HEOM in Wigner space, this approach is effective for identifying purely quantum effects \cite{tanimura1992jcp,sakurai2011jpca}.

For the density matrix $\hat{\bm{\rho }}(t)$, given in Eq.~\eqref{eq:system-densitymatrix}, the Wigner distribution is defined by \cite{tanimura1994jcp,tanimura1997jcp,maruyama1998cpl}
\begin{align}
  \hat{\bm{W}}(p,q;t)&=\sum _{j,k}|j\rangle W_{jk}(p,q;t)\langle k|,
  \label{eq:system-Wigner}
\end{align}
where
\begin{align}
  W_{jk}(p,q;t)&\equiv \frac{1}{2\pi \hbar }\int \mathrm{d}re^{-ipr/\hbar }\rho _{jk}\left(q+\frac{r}{2},q-\frac{r}{2};t\right).
\end{align}
In terms of the Wigner distribution, the $jk$th element of the quantum Liouvillian for the system, $\hat{\bm{\mathcal{L}}}_{\mathrm{qm}}\hat{\bm{\rho }}(q,{q}';t)\equiv (i/\hbar )[\hat{\bm{H}}_{\mathrm{S}},\hat{\bm{\rho }}(q,{q}';t)]$, takes the form
\begin{widetext}
  \begin{align}
    \begin{split}
      \left(\hat{\bm{\mathcal{L}}}_{\mathrm{qm}}\hat{\bm{W}}(p,q;t)\right)_{jk}
      &=\hat{\mathcal{K}}W_{jk}(p,q;t)\\
      &\quad +\frac{i}{\hbar }\sum _{l}\Bigl[U_{jl}(p,q)\ast W_{lk}(p,q;t)-U_{lk}^{\ast }(p,q)\ast W_{jl}(p,q;t)\Bigr],
    \end{split}
    \label{eq:quantum-liouvillian}
  \end{align}  
\end{widetext}
where
\begin{align}
  \hat{\mathcal{K}}&\equiv \frac{p}{m}\frac{\partial }{\partial q}.
\end{align}
Here, $U_{jk}$ are the potentials in Wigner space,
\begin{align}
  U_{jk}(p,q)&\equiv \frac{1}{2\pi \hbar }\int \mathrm{d}re^{-ipr/\hbar }U_{jk}\left(q+\frac{r}{2}\right),
\end{align}
and $\ast $ is the convolution operator, defined as
\begin{align}
  f(p,q)\ast W(p,q)=\int \mathrm{d}p'f(p-p',q)W(p',q).
\end{align}

With the above preparation, the equations of motion for the multi-state (MS) system can be expressed in the form of MS-QHFPE as
\begin{widetext}
  \begin{align}
    \begin{split}
      \frac{\partial }{\partial t}\hat{\bm{W}}^{(\bm{n})}(p,q;t)&=
      -\left\{\hat{\bm{\mathcal{L}}}_{\mathrm{qm}}+\sum _{k}n_{k}\gamma _{k}\bm{I}-c_{\delta }\hat{\bm{\Phi }}^{2}\right\}\hat{\bm{W}}^{(\bm{n})}(p,q;t)\\
      &\quad -\sum _{k}\hat{\bm{\Phi }}\hat{\bm{W}}^{(\bm{n}+\bm{e}_{k})}(p,q;t)\\
      &\quad -\sum _{\bm{m}\leq \bm{n},\left|\bm{m}\right|\geq 1}\frac{\bm{n}!}{\bm{m}!(\bm{n}-\bm{m})!}\bm{\gamma }^{\bm{m}}\bm{r}^{\bm{m}}\hat{\bm{\Lambda }}^{(\left|\bm{m}\right|)}\hat{\bm{W}}^{(\bm{n}-\bm{m})}(p,q;t)\\
      &\quad -\sum _{k}n_{k}\gamma _{k}c_{k}\bm{\hat{\Phi }}\hat{\bm{W}}^{(\bm{n}-\bm{e}_{k})}(p,q;t),
    \end{split}
    \label{eq:heom}
  \end{align}
\end{widetext}
where $\bm{n}=(n_{0},\dots ,n_{K})$ is a $(K+1)$-dimensional vector whose components are all non-negative integers, and $\bm{e}_{k}$ is the $k$th unit vector.
We have introduced $\bm{\gamma }\equiv (\gamma _{0},\dots ,\gamma _{K})$, $\bm{r}\equiv (r_{0},\dots ,r_{K})$, and the following multi-index notation:
$\bm{n}\leq \bm{m}\equiv \bigwedge _{k}n_{k}\leq m_{k}$, $\left|\bm{n}\right|\equiv \sum _{k}n_{k}$, $\bm{n}!\equiv \prod _{k}n_{k}!$ and $\bm{\gamma }^{\bm{n}}\equiv \prod _{k}\gamma _{k}^{n_{k}}$.
The vector $\bm{n}$ represents the index of the hierarchy.
In this formalism, only the first element in the hierarchy, $\hat{\bm{W}}^{(0,\dots ,0)}(p,q;t)\equiv \hat{\bm{W}}(p,q;t)$ has a physical meaning.
The rest of the elements serve only to allow treatment of non-perturbative, non-Markovian system-bath interaction \cite{tanimura2014jcp,tanimura2015jcp}.
The operators $\hat{\bm{\Phi }}$ and $\hat{\bm{\Lambda }}^{(l)}$ describe the interaction between the system and the bath.
They are defined as
\begin{align}
  \hat{\bm{\Phi }}&\equiv \bm{I}\otimes \frac{i}{\hbar }\Bigl(V(p,q)-V^{\ast }(p,q)\Bigr)\ast 
  \intertext{and}
  \hat{\bm{\Lambda }}^{(l)}&\equiv (\bm{\hat{\Psi }}^{\times })^{l}\bm{I}\otimes \hat{\mathcal{K}},
\end{align}
where
\begin{align}
  \hat{\bm{\Psi }}&\equiv \bm{I}\otimes \frac{1}{2}\Bigl(V(p,q)+V^{\ast }(p,q)\Bigr)\ast ,
\end{align}
and the commutation hyper-operator $\times $ is defined as $\hat{\bm{A}}^{\times }\hat{\bm{B}}\equiv [\hat{\bm{A}},\hat{\bm{B}}]$.
Here $V(p,q)$ is defined as
\begin{align}
  V(p,q)\equiv \frac{1}{2\pi \hbar }\int \mathrm{d}re^{-ipr/\hbar }V\left(q+\frac{r}{2}\right).
\end{align}
Note that $\hat{\bm{\Lambda }}^{(l)}=0$ for all $l\geq 3$, because $\hat{\mathcal{K}}$ is a second-order differential operator in the coordinate space, and $(V(q)^{\times })^{3}\hat{\mathcal{K}}$ vanishes.

Here we restrict our analysis to the case of linear interaction, $V(q)=q$.
For this reason, our treatment describes vibrational relaxation only.
In this case, the operators are expressed as
\begin{align}
  \hat{\bm{\Phi }}&=-\bm{I}\otimes \frac{\partial }{\partial p},\\
  \hat{\bm{\Psi }}&=\bm{I}\otimes q,\\
  \hat{\bm{\Lambda }}^{(1)}&=-\bm{I}\otimes \frac{p}{m},\\
  \intertext{and}
  \hat{\bm{\Lambda }}^{(2)}&=\bm{0}.
\end{align}
In this study, we consider the high temperature case, characterized by the condition $\coth (\beta \hbar \gamma /2)\approx 2/\beta \hbar \gamma $.
In this case, the right-hand sides of Eqs.~\eqref{eq:expansion-of-bath-relaxation} and \eqref{eq:expansion-of-bath-correlation} reduce to single exponential functions (i.e.
$K=0$) and we have
\begin{align}
  \Psi _{\mathrm{B}}(t)&=r_{0}\cdot \gamma e^{-\gamma \left|t\right|}=m\zeta \cdot \gamma e^{-\gamma _{k}\left|t\right|}\\
  \label{eq:psi-Drude}
  \intertext{and}
  C_{\mathrm{B}}(t)&=c_{0}\cdot \gamma e^{-\gamma \left|t\right|}=\frac{m\zeta }{\beta }\cdot \gamma e^{-\gamma \left|t\right|}
\end{align}

Note that Eq.~\eqref{eq:heom} consists of an infinite number of simultaneous equations.
In order to make these equations tractable, we introduce a ``terminator'' for the $N$th equation, where $N$ is the depth of hierarchy \cite{tanimura1990pra,ishizaki2005jpsj,tanimura2006jpsj,tanimura2014jcp,tanimura2015jcp,tanimura1991pra,tanimura1992jcp}.
We chose $N$ such that $(N+1)\gg \omega _{\mathrm{c}}/\gamma $, where $\omega _{\mathrm{c}}$ is the characteristic frequency of the system.
Here, we employ the equation
\begin{widetext}
  \begin{align}
    \begin{split}
      \frac{\partial }{\partial t}\hat{\bm{W}}^{(N)}(p,q;t)&=
      -\left\{\hat{\bm{\mathcal{L}}}_{\mathrm{qm}}+n\gamma \bm{I}-\hat{\bm{\Phi }}(r_{0}\hat{\bm{\Lambda }}^{(1)}+c_{0}\bm{\hat{\Phi }})\right\}\hat{\bm{W}}^{(N)}(p,q;t)\\
      &\quad -n\gamma (r_{0}\hat{\bm{\Lambda }}^{(1)}+c_{0}\bm{\hat{\Phi }})\hat{\bm{W}}^{(N-1)}(p,q;t).
    \end{split}
    \label{eq:GMFP} 
  \end{align}
\end{widetext}
Note that in the Markovian limit, $\gamma \gg \omega _{\mathrm{c}}$, we can set $N=0$.
In this case, Eq.~\eqref{eq:GMFP} reduces to the multi-state quantum Fokker-Planck equation \cite{tanimura1994jcp}
\begin{align}
  \begin{split}
    &\frac{\partial }{\partial t}\hat{\bm{W}}^{(0)}(p,q;t)\\
    &=-\left\{\hat{\bm{\mathcal{L}}}_{\mathrm{qm}}-\zeta \frac{\partial }{\partial p}\left(p+\frac{m}{\beta }\frac{\partial }{\partial p}\right)\otimes \bm{I}\right\}
    \hat{\bm{W}}^{(0)}(p,q;t).
  \end{split}
  \label{eq:heomW}
\end{align}  

\section{Numerical results}
\label{sec:optical_observables}
With the HEOM formalism, it is possible to calculate nonlinear response functions \cite{tanimura2000jpsj,ishizaki2006jcp,sakurai2011jpca,tanimura2006jpsj,ikeda2015jcp,ito2016jcp,tanimura2014jcp,tanimura2015jcp,tanimura2015jcp,tanimura1994jcp,tanimura1997jcp,maruyama1998cpl,chen2010jcp,chen2011jcp,hein2012njp,kreisbeck2013jpcb,tanimura1994jcp,tanimura2012jcp,gelin2013jcp,dijkstra2015jcp}.
The MS-QHFP approach used in this work can be applied to systems with potentials of arbitrary form for the calculation of linear and nonlinear spectra.
In the present study, we computed linear absorption (1D) and two-dimensional (2D) electronic spectra for a displaced anharmonic system, whose response functions cannot be obtained analytically, unlike in the harmonic case.
Moreover, this method can be used to treat diabatic coupling that depends nonlinearly on the coordinate.
This is necessary to describe the large structural changes undergone by the system in photoisomerization.

The first-order and third-order response functions are expressed as \cite{tanimura2000jpsj,ishizaki2006jcp,sakurai2011jpca,ikeda2015jcp,ito2016jcp}
\begin{widetext}
  \begin{align}
    R^{(1)}(t)&=\left(\frac{i}{\hbar }\right)\mathrm{Tr}\left\{\hat{\bm{\mu }}\bm{\mathcal{G}}(t)\hat{\bm{\mu }}_{W}^{\times }\hat{\bm{W}}_{\mathrm{eq}}\right\}
    \label{eq:R1}\\
    \intertext{and}
    R^{(3)}(t_{3},t_{2},t_{1})&=\left(\frac{i}{\hbar }\right)^{3}\mathrm{Tr}\left\{\hat{\bm{\mu }}\bm{\mathcal{G}}(t_{3})\hat{\bm{\mu }}^{\times }_{W}\bm{\mathcal{G}}(t_{2})\hat{\bm{\mu }}^{\times }_{W}\bm{\mathcal{G}}(t_{1})\hat{\bm{\mu }}^{\times }_{W}\hat{\bm{W}}_{\mathrm{eq}}\right\}.
    \label{eq:R3}
  \end{align}
\end{widetext}
Here, $\hat{\bm{\mu }}$ is the dipole operator of the system and $\bm{\mathcal{G}}(t)$ is the Green's function in the absence of a laser interaction evaluated from Eq.~\eqref{eq:heom}.
The operator $\hat{\bm{\mu }}^{\times }_{W}\equiv \hat{\bm{\mu }}^{\rightarrow }_{W}-\hat{\bm{\mu }}_{W}^{\leftarrow }$ is the commutator of the dipole operator, where $\hat{\bm{\mu }}_{W}^{\rightarrow }\hat{\bm{W}}(t)$ and $\hat{\bm{\mu }}_{W}^{\leftarrow }\hat{\bm{W}}(t)$ are the partial Wigner representations of $\hat{\bm{\mu }}\hat{\bm{\rho }}(t)$ and $\hat{\bm{\rho }}(t)\hat{\bm{\mu }}$, and they are described by
\begin{align}
  (\hat{\bm{\mu }}^{\rightarrow }_{W}\hat{\bm{W}})_{jk}(p,q;t)&=\sum _{l}\mu _{jl}(p,q)\ast W_{lk}(p,q;t)\\
  \intertext{and}
  (\hat{\bm{\mu }}^{\leftarrow }_{W}\hat{\bm{W}})_{jk}(p,q;t)&=\sum _{l}\mu _{lk}^{\ast }(p,q)\ast W_{jl}(p,q;t),
\end{align}
respectively, where $\mu _{jk}(p,q)\equiv \frac{1}{2\pi \hbar }\int \mathrm{d}re^{-ipr/\hbar }\mu _{jk}(q+r/2)$ .
To evaluate these response functions, we developed a computational program incorporating MS-QHFPE presented in Sec.
II.
We first ran the computational program to evaluate Eq.~\eqref{eq:R1} or \eqref{eq:R3} for a sufficiently long time to obtain a true thermal equilibrium state.
The full hierarchy members of $W_{00}^{(\bm{n})}(p,q; 0)$ were then used to obtain the correlated initial thermal equilibrium state.
The system was excited by the first interaction, $\hat{\bm{\mu }}^{\times }_{W}$, at $t=0$.
The evolution of the perturbed elements were then computed by running the program for the MS-QHFP up to some time $t_{1}$.
The linear response function defined in Eq.~\eqref{eq:R1} was then calculated from the expectation value of $\hat{\bm{\mu }}$.
We carried out such calculations in both the first-order case, employing $R^{(1)}(t)$ and in the third-order case, employing $R^{(3)}(t_{3},t_{2}, t_{1})$.
The third-order case, we calculated $R^{(3)}(t_{3},t_{2}, t_{1})$ for various values of $t_{1}$, $t_{2}$ and $t_{3}$ by extending the method employed in the first-order case.
The 1D spectrum and the 2D correlation spectrum are evaluated as \cite{gallagher1999jpca,ge2002jpca}
\begin{align} 
  I^{\mathrm{(abs)}}(\omega _{1})&=\omega \mathrm{Im}\int _{0}^{\infty }\mathrm{d}t_{1}e^{i\omega _{1}t_{1}}R^{(1)}(t_{1})
  \label{eq:I1}
  \intertext{and}
  I^{\mathrm{(corr)}}(\omega _{3},t_{2},\omega _{1})
  &= I^{\mathrm{(NR)}}(\omega _{3},t_{2},\omega _{1}) + I^{\mathrm{(R)}}(\omega _{3},t_{2},\omega _{1}),
  \label{eq:I3}
\end{align}
where the non-rephasing and rephasing parts of the signal are defined by
\begin{widetext}
  \begin{align} 
    I^{\mathrm{(NR)}}(\omega _{3},t_{2},\omega _{1}) &=
    \mathrm{Im}\int _{0}^{\infty }\mathrm{d}t_{3}\int _{0}^{\infty }\mathrm{d}t_{1}e^{i\omega _{3}t_{3}}e^{i\omega _{1}t_{1}}R^{(3)}(t_{3},t_{2},t_{1}),
    \label{eq:nonreph}
    \intertext{and}
    I^{\mathrm{(R)}}(\omega _{3},t_{2},\omega _{1}) &=
    \mathrm{Im}\int _{0}^{\infty }\mathrm{d}t_{3}\int _{0}^{\infty }\mathrm{d}t_{1}e^{i\omega_{3}t_{3}}e^{-i\omega _{1}t_{1}}R^{(3)}(t_{3},t_{2},t_{1}).
    \label{eq:reph}
  \end{align}
\end{widetext}
The transient absorption spectrum is expressed in terms of the third-order response function as
\begin{align}
  I^{\mathrm{(tas)}}(\omega ,T)&=\omega \mathrm{Im}\int _{0}^{\infty }\mathrm{d}te^{i\omega t}R^{(3)}(t,T,0).
  \label{eq:transient-absorption}
\end{align}

\begin{SCfigure*}
  \includegraphics[scale=0.7]{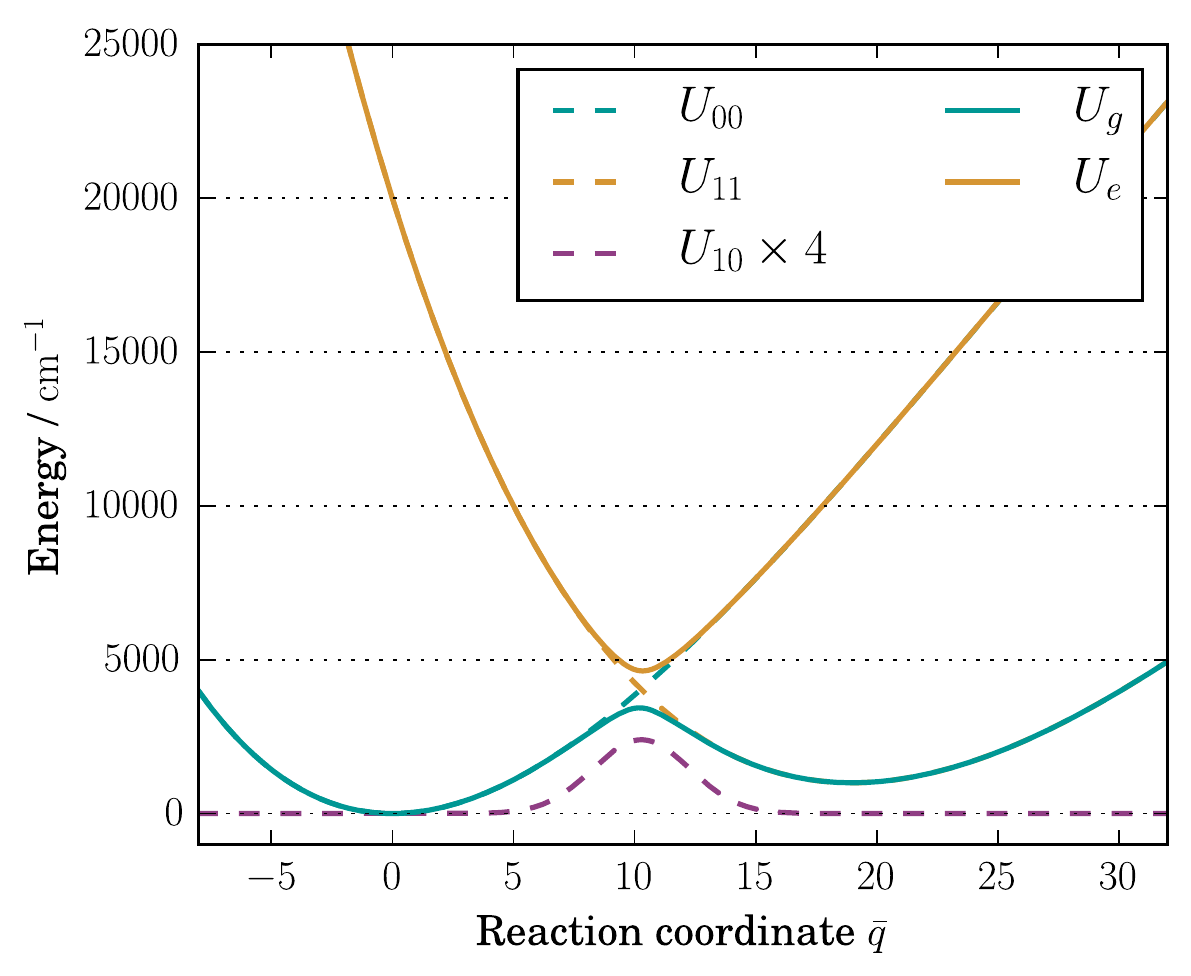}
  \caption{ Diabatic and adiabatic electronic potentials.
    The dashed orange and dashed green curves represent the ground and excited diabatic potentials, while the dashed purple curve represents the diabatic coupling.
    The solid green and solid orange curves represent the ground and excited adiabatic potential surfaces, respectively.}
  \label{fig:pot}
\end{SCfigure*}
We applied the MS-QHFPE approach to a two-level system, The levels $0$ and $1$, represent the reactant and product states, respectively.
Hereafter, we employ the dimensionless coordinate and momentum defined by $\bar{q}\equiv (m\omega _{0}/\hbar )^{1/2}q$ and $\bar{p}\equiv (m\hbar \omega _{0})^{-1/2}p$, where $\omega _{0}$ is the characteristic frequency of the system.
The diabatic potential surfaces are expressed as displaced Morse potentials:
\begin{align}
  U_{00}(\bar{q})&\equiv D_{0}\left(1-e^{\sqrt {\omega _{0}/2D_{0}\mathstrut }(\bar{q}-\bar{q}_{0})}\right)^{2},\\
  U_{11}(\bar{q})&\equiv D_{1}\left(1-e^{\sqrt {\mathstrut \omega _{1}/2D_{1}}(\bar{q}-\bar{q}_{1})}\right)^{2} + \Delta E.
\end{align}
Here, $D_{j}$ and $\omega _{j}$ are the dissociation energy and vibrational frequency at the minimum of each potential in the $j$th state, and $\Delta E$ is the difference between the energies of the product and reactant states.

We assume that the diabatic coupling possesses the Gaussian form 
\begin{align}
  U_{10}(\bar{q})=U_{01}(\bar{q})&=\Delta Ve^{-(\bar{q}-\bar{q}^{\ast })^{2}/2\bar{\sigma }^{2}}.
\end{align}
We set $\omega _{0}=100~\mathrm{cm}^{-1}$ and $\omega _{1}=80~\mathrm{cm}^{-1}$, which are typical values for the intramolecular motion of large molecules.
The other parameters were set as follows:
$D_{0}=70,000~\mathrm{cm}^{-1}$, $D_{1}=50,000~\mathrm{cm}^{-1}$, $\Delta E=1,000~\mathrm{cm}^{-1}$, $\Delta V=600~\mathrm{cm}^{-1}$, $\bar{q}_{0}=0.00$, $\bar{q}_{1}=18.98$, $\bar{q}^{\ast }=10.28$, and $\bar{\sigma }=2.0$.
With these values, the electronic resonant frequencies at the stable point of each diabatic state are $\nu _{0}\equiv [U_{11}(\bar{q}_{0})-U_{00}(\bar{q}_{0})]/\hbar =20,000~\mathrm{cm}^{-1}$ and $\nu _{1}\equiv [U_{00}(\bar{q}_{1})-U_{11}(\bar{q}_{1})]/\hbar =10,083~\mathrm{cm}^{-1}$.
The adiabatic potential functions are given by
\begin{widetext}
  \begin{align}
    U_{g}(\bar{q})&=\frac{U_{11}(\bar{q})+U_{00}(\bar{q})}{2}-\sqrt {\mathstrut \left(\frac{U_{11}(\bar{q})-U_{00}(\bar{q})}{2}\right)^{2}+U_{10}(\bar{q})^{2}}\\
    \intertext{and}
    U_{e}(\bar{q})&=\frac{U_{11}(\bar{q})+U_{00}(\bar{q})}{2}+\sqrt {\mathstrut \left(\frac{U_{11}(\bar{q})-U_{00}(\bar{q})}{2}\right)^{2}+U_{10}(\bar{q})^{2}}.
  \end{align}
\end{widetext}
The diabatic (dashed curve) and adiabatic (solid curve) potential curves are depicted in Fig.~\ref{fig:pot}.
We adopt the Condon approximation and assume that the dipole operator takes the form $\hat{\bm{\mu }}=\mu _{0}(|0\rangle \langle 1|+|1\rangle \langle 0|)$, where $\mu _{0}$ is a constant, independent of the coordinate.
Note that, if we wish to separate the two-dimensional signal into the rephasing and non-rephasing parts using the Liouville path based approach \cite{tanimura2012jcp,ishizaki2006jcp}, we must define rephasing and non-rephasing Liouville pathways in the adiabatic basis instead of the diabetic basis to maintain a proper phase associated with each Liouville path that involves photoisomerization.
In the present MS-QHFP approach, we can include a non-Condon dipolar interaction \cite{tanimura1993pre} without difficulty \cite{tanimura2000jpsj,sakurai2011jpca,ikeda2015jcp,ito2016jcp}.
Such an interaction is necessary to calculate transient impulsive Raman spectra \cite{takeuchi2008science}.

\begin{SCfigure*}
  \includegraphics[scale=0.6]{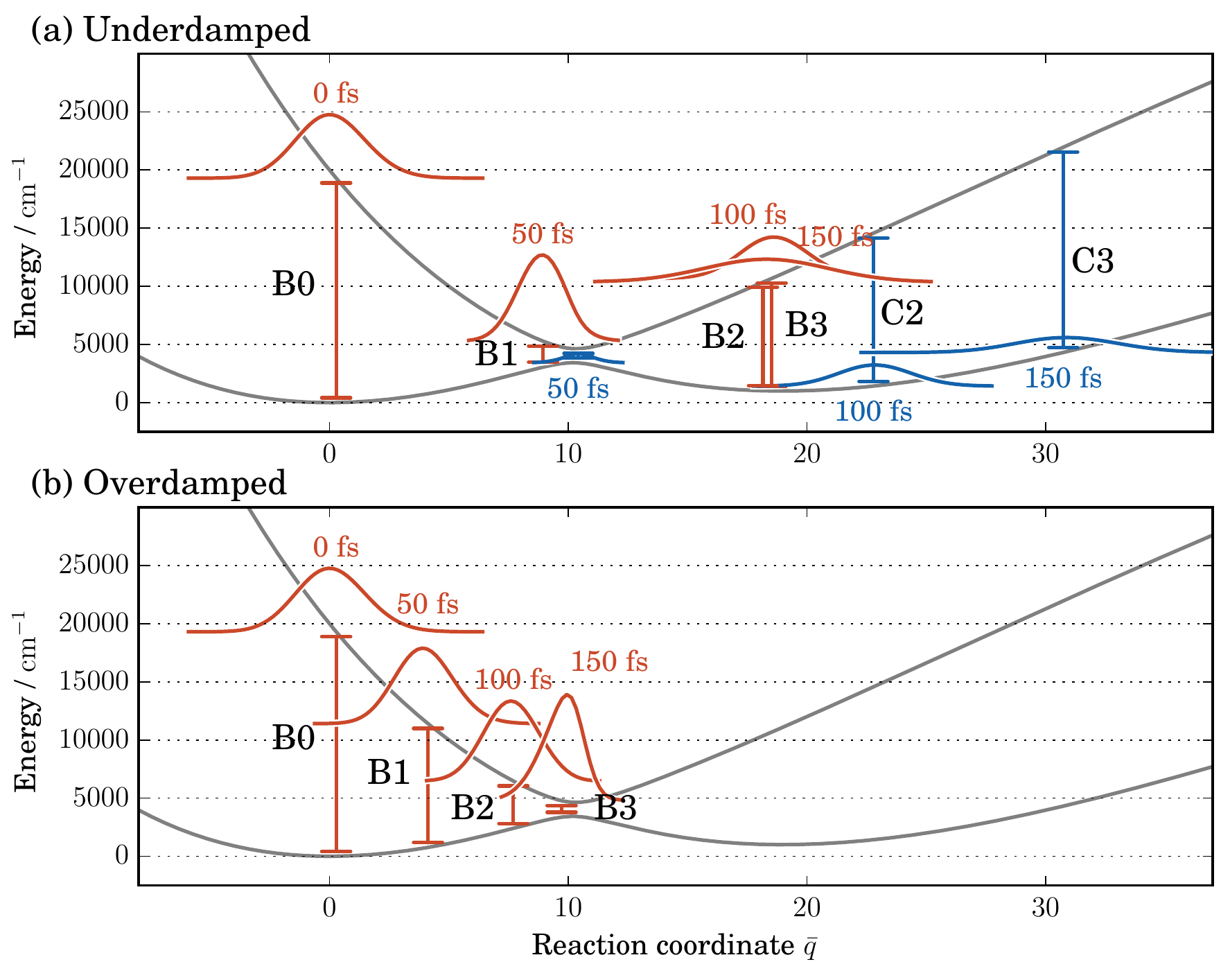}
  \caption{ Snapshots of excited wavepackets in the adiabatic basis calculated under (a) underdamped and (b) overdamped conditions for waiting times $t_{2}=0~\mathrm{fs}$, $50~\mathrm{fs}$, $100~\mathrm{fs}$, and $150~\mathrm{fs}$.
    The red and blue curves represent the distributions for the excited and ground states, respectively.
    The labels B0, B1, B2, B3, C2, and C3 correspond to the peaks labeled in Fig.~\ref{fig:multi_corr}. }
  \label{fig:wd}
\end{SCfigure*}
\begin{figure*}
  \centering
  \includegraphics[scale=0.5]{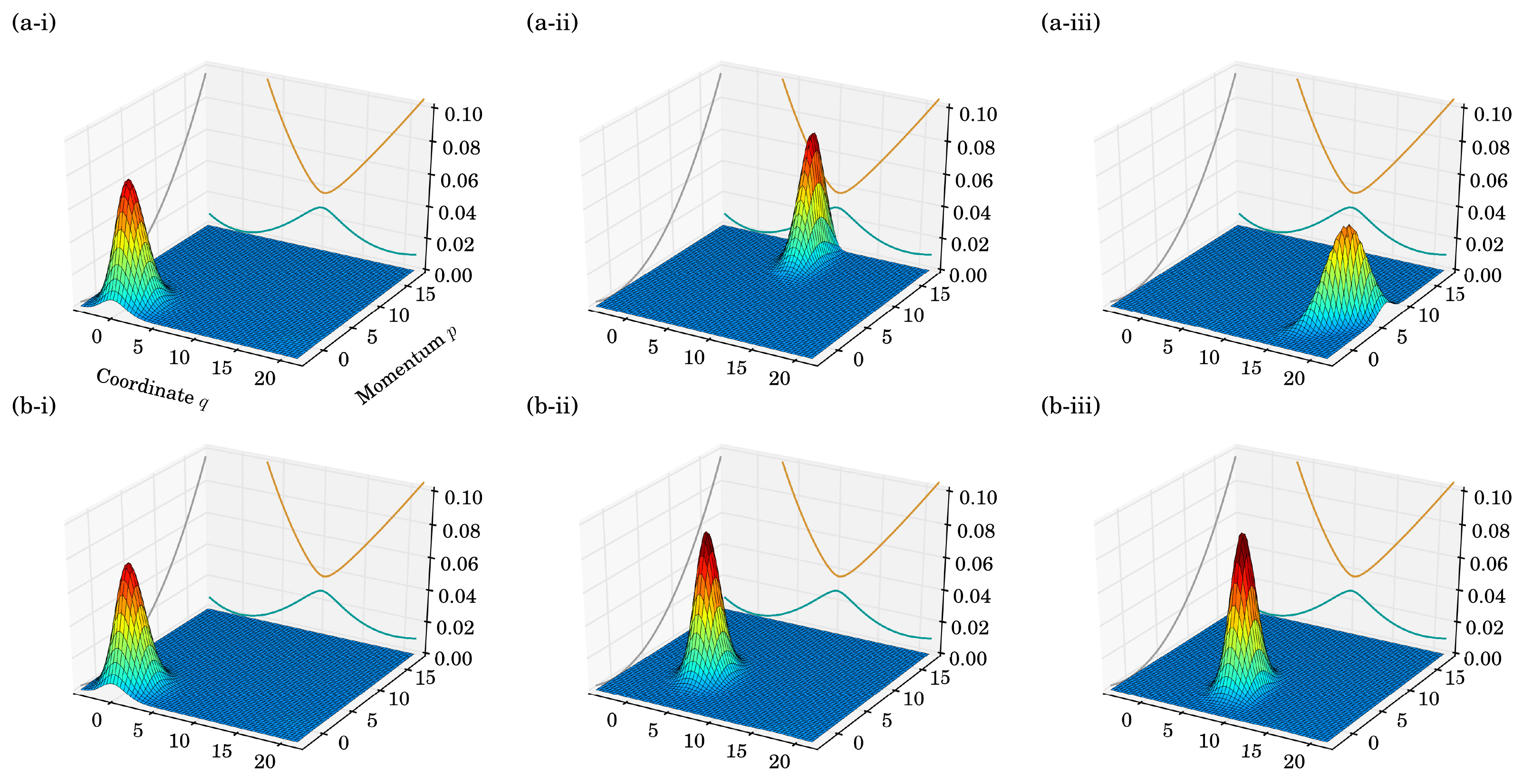}
  \caption{Snapshots of wavepackets for an adiabatic electronic excited state calculated under (a) underdamped and (b) overdamped conditions with three values of the waiting times, (i) $0~\mathrm{fs}$, (ii) $50~\mathrm{fs}$, (iii) $100~\mathrm{fs}$.
  }
  \label{fig:W}
\end{figure*}
We used the linear system-bath coupling function $V(q)=q$ and the coupling strengths and cutoff frequencies were chosen to realize underdamped and overdamped conditions, with $\zeta =40~\mathrm{cm}^{-1}$ and $\gamma =500~\mathrm{cm}^{-1}$ in the former case, and $\zeta =400~\mathrm{cm}^{-1}$ and $\gamma =\infty $ in the latter.
Although we cannot apply high-temperature approximation to the latter case, we use Eq.~\eqref{eq:heomW} because it behaves semi-classically under fast and strong damping condition.
The bath temperature was set to $T = 300~\mathrm{K}$ ($\beta \hbar \omega _{0} \approx 0.48$).
The numerical calculations carried out to integrate Eq.~\eqref{eq:heom} were performed using the fourth-order exponential integrator method \cite{almohy2011sjsc,wilkins2015jctc,tsuchimoto2015jctc}, with a discrete mesh expression in Wigner space.
The mesh size was set to $N_{\bar{p}}\times N_{\bar{q}}=128\times 512$, with mesh ranges $-25\leq \bar{p}\leq 25$ and $-20\leq \bar{q}\leq 80$.

We evaluated the kinetic term in the quantum Liouvillian, given in Eq.~\eqref{eq:quantum-liouvillian}, using the fifth-order upwind difference scheme that is expressed as
\begin{widetext}
  \begin{align}
    \frac{\partial W(p_{k},q_{j})}{\partial q}&=
    \left\{
    \begin{matrix}
      &\begin{split}
        &\frac{1}{60\Delta _{q}}\Bigl(2W(p_{k},q_{j+3})-15W(p_{k},q_{j+2})+60W(p_{k},q_{j+1})\\
        &\quad \quad \quad \quad -20W(p_{k},q_{j})-30W(p_{k},q_{j-1})+3W(p_{k},q_{j-2})\Bigr)
      \end{split}
      &(p_{k}<0)\\
      &\begin{split}
        &\frac{1}{60\Delta _{q}}\Bigl(-3W(p_{k},q_{j+2})+30W(p_{k},q_{j+1})+20W(p_{k},q_{j})\\
        &\quad \quad \quad \quad -60W(p_{k},q_{j-1})+15W(p_{k},q_{j-2})-2W(p_{k},q_{j-3})\Bigr)
      \end{split}
      &(p_{k}\geq 0).\\
    \end{matrix}
    \right.
  \end{align}
\end{widetext}
The potential terms in Eq.~\eqref{eq:quantum-liouvillian} were evaluated using a fast Fourier transform (FFT) that was derived by expanding a discrete formulation of the Liouvillian in Wigner space \cite{kim2007jap} for multi-state.
We implemented this calculation on a graphic processor unit (GPU) using the FFT routine in the CUDA library (cuFFT).
The depth of the hierarchy was chosen to be $N=5$ in the underdamped case and $N=0$ in the overdamped case.

\subsection{Wavepacket dynamics}
First we illustrate snapshots of excited wavepackets created by a pair of impulsive pump pulses set by $t_{1}=0$.
These wavepackets are written
\begin{align}
  P_{jk}^{\mathrm{(ad)}}(q;t)&\equiv \int \mathrm{d}pW_{jk}^{\mathrm{(ad)}}(p,q;t),
\end{align}
where $\hat{\bm{W}}^{\mathrm{(ad)}}$ is the Wigner distribution in the adiabatic basis evaluated using the Wigner distribution in the diabatic basis, defined as
\begin{align}
    \hat{\bm{W}}(p,q;t_{2})&\equiv -\left(\frac{i}{\hbar }\right)^{2}\mathcal{G}(t_{2})\hat{\bm{\mu }}_{W}^{\rightarrow }\hat{\bm{\mu }}_{W}^{\leftarrow }\hat{\bm{W}}_{\mathrm{eq}}.
\end{align}
In Fig.~\ref{fig:wd}, we present $P_{11}^{\mathrm{(ad)}}(q;t)$ and $P_{00}^{\mathrm{(ad)}} (q;t)$ in the underdamped and overdamped cases for several values of $t_{2}$.
The Wigner distribution $W_{11}^{\mathrm{(ad)}}(p,q;t)$, which was used for the calculation of $P_{11}^{\mathrm{(ad)}} (q;t)$, is presented in Fig.~\ref{fig:W}.
As seen in Fig.~\ref{fig:wd}, the excitation, relaxation, and nonadiabatic transition processes, followed by an impulsive excitation, are clearly described.
While the distribution function $P_{11}^{\mathrm{(ad)}} (q;t)$ merely represents the location of a wavepacket, the Wigner distribution $W_{11}^{\mathrm{(ad)}} (p,q;t)$ contains all of the information for the dynamics.
In the underdamped case depicted in Fig.~\ref{fig:W}(a), the excited wavepacket passes through a crossing region ($\bar{q}\sim 10$) maintaining a large momentum, and as a results, the adiabatic approximation breaks down.
In the overdamped case depicted in Fig.~\ref{fig:W}(b), however, the momentum decays rapidly and for this reason, the adiabatic approximation holds well.
It is thus seen that using the distribution function in phase space, we can analyze the behavior of nonadiabatic processes more completely than if we use only the distribution function in coordinate space.

\subsection{One-dimensional spectrum}
\begin{figure}
  \centering
  \includegraphics[scale=0.6]{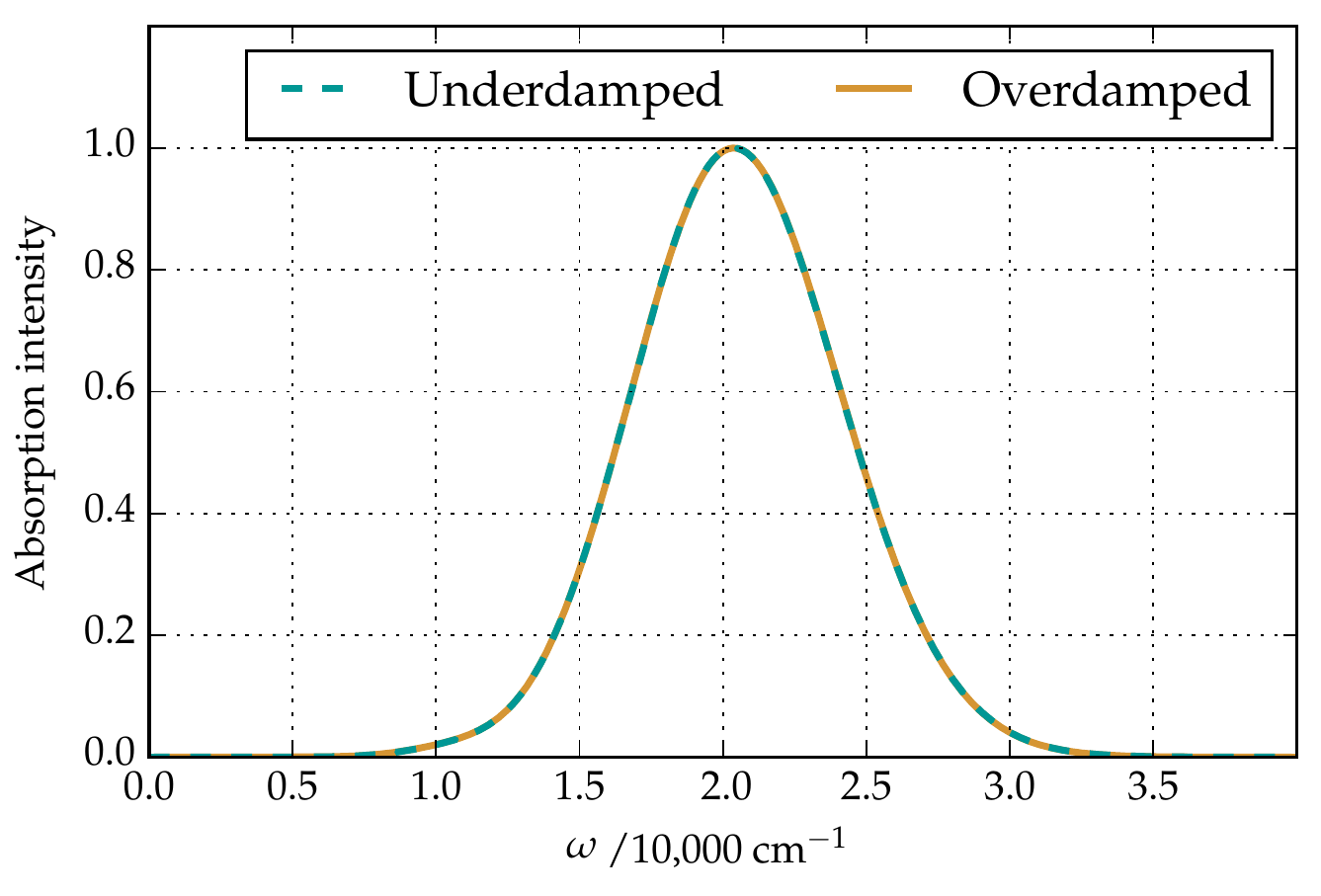}
  \caption{Linear absorption spectra, $I^{\mathrm{(abs)}}(\omega )$, calculated under underdamped (green dashed curved) and overdamped (orange solid curve) conditions.
    The two curves are nearly identical, because linear absorption spectra are insensitive to the dissipative effects of the wavepacket dynamics.}
  \label{fig:linear}
\end{figure}
In Fig.~\ref{fig:linear}, 1D spectra are presented for the underdamped (green dashed curve) and overdamped (orange solid curve) cases.
In the situation considered here, the displacement of the Morse oscillators is so large that the transition occurs mainly between the ground state and near continuum dissociation states, in which the wavepacket does not exhibit coherent oscillations.
Thus, the profile of a signal is determined by the shape of the ground state wavepacket.
As a result, the profiles of the spectra in the weak and strong damping cases are almost identical \cite{tanimura1997jcp}.

\subsection{Two-dimensional correlation spectrum}
\begin{figure*}
  \centering
  \includegraphics[scale=0.7]{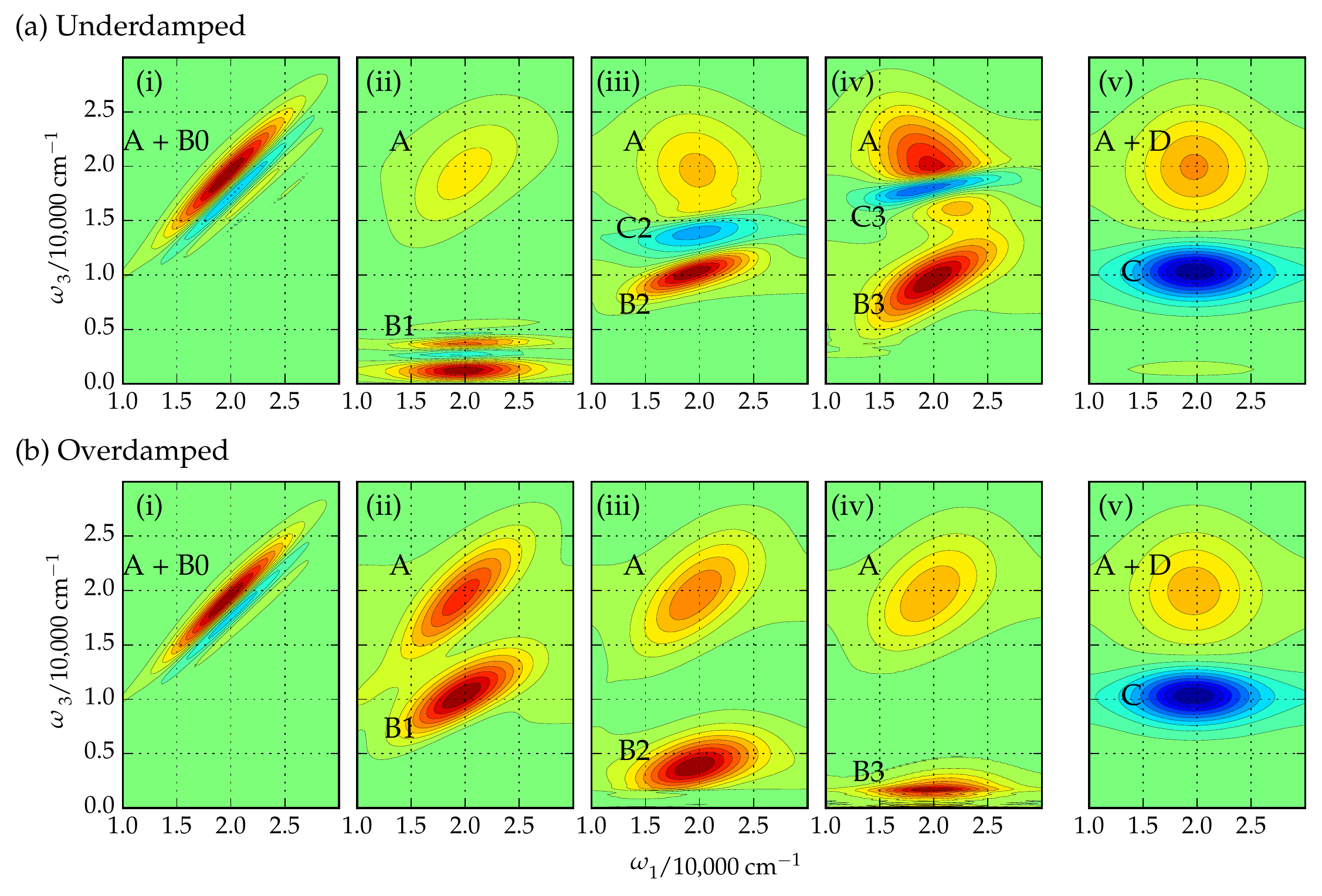}
  \caption{Two-dimensional correlation spectra, $I^{\mathrm{(corr)}}(\omega _{3},t_{2},\omega _{1})$, calculated under (a) underdamped and (b) overdamped dissipative conditions with various values of the waiting times:
    (i) $t_{2}=0~\mathrm{fs}$, (ii) $t_{2}=50~\mathrm{fs}$, (iii) $t_{2}=100~\mathrm{fs}$, (iv) $t_{2}=150~\mathrm{fs}$, (v) $t_{2}=5,000~\mathrm{fs}$.
    The red and blue areas represent emission and absorption, respectively.
    The intensities are normalized with respect to the maximum peak value of each spectrum, because we are interested in the profile of each spectrum. }
  \label{fig:multi_corr}
\end{figure*}
The calculated 2D correlation spectra for the (a) underdamped and (b) overdamped cases are presented in Fig.~\ref{fig:multi_corr}.
The peaks labeled ``A'' correspond to the transition from the electronic ground state (hole contribution or bleaching), while the peaks labeled ``B2'' are from the transition from the electronic excitation state (particle contribution or stimulated emission).
The positions of the peaks labeled ``B1'' vary as functions of $t_{2}$, as seen in Figs.~\ref{fig:wd} (a) and (b).
At $t_{2}=0$, positive and negative elongated diagonal peaks centered near $(\omega _{1},\omega _{3})=(20,000~\mathrm{cm}^{-1}, 20,000~\mathrm{cm}^{-1})$ are observed in both cases of the 2D correlation spectrum.
This phenomenon, which we call a ``coherent fringe'', is due to the fast motion of the excited wavepackets in the period $t=[0,t_{1}]$.
By analyzing the positions of these elongated peaks, we can estimate the gradient of the excited state potential (see Appendix \ref{sec:coherent_fringe}).
At $t_{2}=100~\mathrm{fs}$ in the underdamped case depicted in Fig.~\ref{fig:multi_corr} (a-iii), the peak labeled ``B2'' moves in the direction of increasing $\omega _{3}$, because the particle wavepacket moves up the potential as $t_{2}$ increases.
The negative (absorption) peaks labeled ``C2'' appear because a part of the particle wavepacket transfers to the ground state (see Fig.~\ref{fig:wd}), which causes absorption from the ground state to the excited state.
In the overdamped case depicted in Fig.~\ref{fig:multi_corr} (b-iii), the particle wavepacket labeled ``B2'' reaches the crossing point.
This peak then decays gradually, due to population relaxation to the metastable or ground state as illustrated in Fig.~\ref{fig:wd} (b).
At $t_{2}=150~\mathrm{fs}$ depicted in Fig.~\ref{fig:wd}(a-iv), the negative elongated peaks labeled ``C3'' moves in the direction of increasing $\omega _{3}$ following the movement of wavepacket in the ground sate as illustrated in Fig.~\ref{fig:wd} (a).
At this time, the particle wavepacket bounces and moves back toward the minimum of the potential.
As a result, the peak labeled ``B3'' in Fig.~\ref{fig:multi_corr} (a-iv) spreads.
In the overdamped case depicted in Fig.~\ref{fig:multi_corr} (b-iv), the motion of the wavepacket is suppressed by the strong relaxation, and the particle contribution slowly decays.
As seen in Figs.~\ref{fig:multi_corr} (a-v) and (b-v), at $t_{2}=5,000~\mathrm{fs}$ the profile of the peaks from the hole and particle in the reactant state labeled ``A+D'' possesses symmetrical circular shape.
This results from the loss of the coherence between the initial ground state and final states.
The negative peaks labeled ``C'' arise from the ground state wavepacket in the product state.

\subsection{Transient absorption spectra}
Next we present transient absorption spectra (TAS), which have been used to investigate photoisomerization process experimentally.
We do this to clarify the difference between 2D electronic spectroscopy and TAS.
The TAS is obtained from the 2D correlation spectrum as $I^{\mathrm{(tas)}}(\omega ,T)=\omega \int \mathrm{d}\omega _{1}I^{\mathrm{(corr)}}(\omega ,T,\omega _{1})$.
\begin{SCfigure*}
  \includegraphics[scale=0.75]{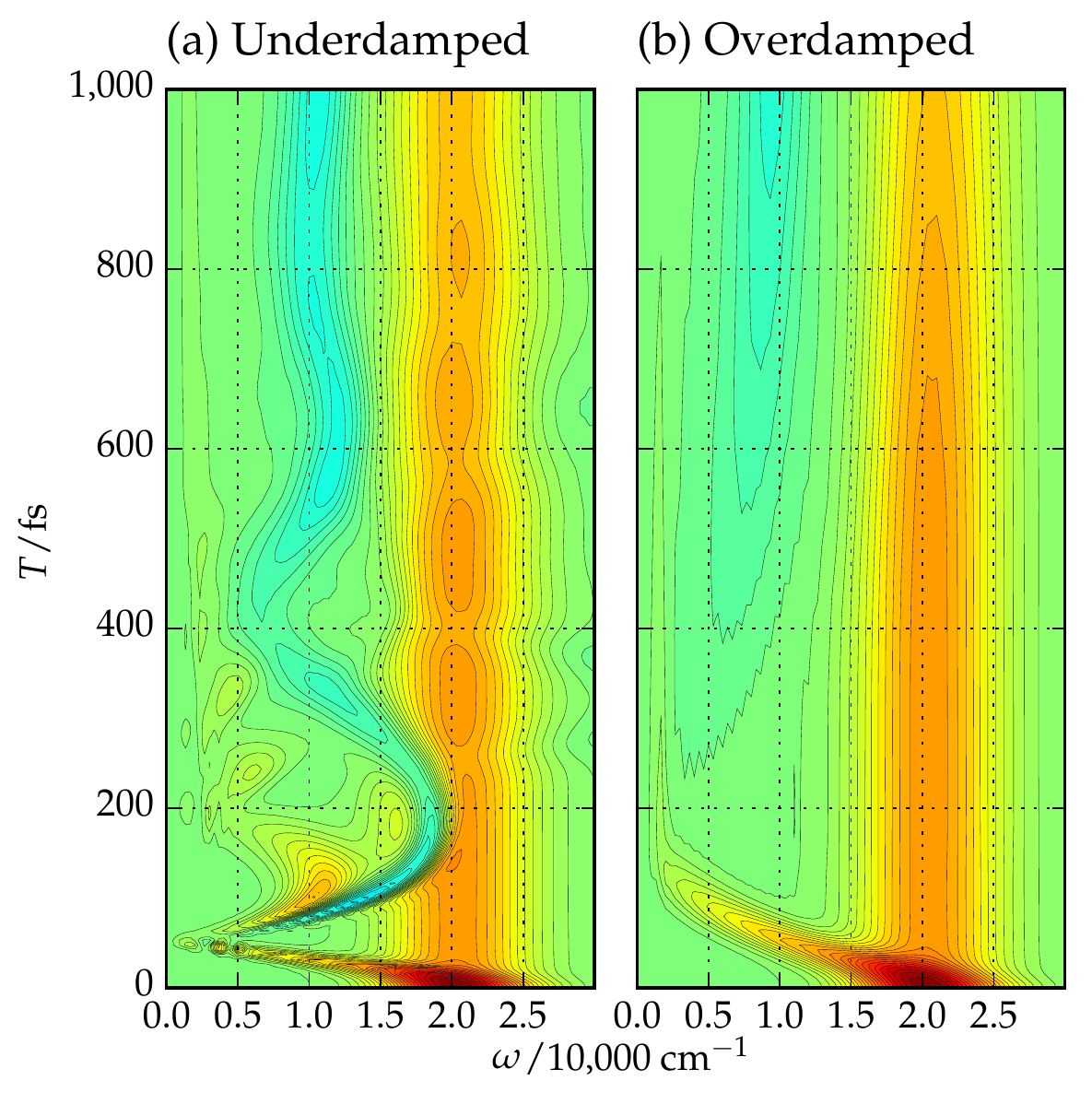}
  \caption{Transient absorption spectra, $I^{\mathrm{(tas)}}(\omega ,T)$, calculated under (a) underdamped and (b) overdamped conditions.
    The red and blue areas represent emission and absorption, respectively. }
  \label{fig:multi_pp}
\end{SCfigure*}
In Fig.~\ref{fig:multi_pp}, a comparison is made between TAS in the underdamped and overdamped cases.
Here, although we can observe the motion of the hole and particle as functions of $t_{2}$, we cannot study the coherence of the pump and probe processes.
Moreover, a coherent fringe arising from the motion of the excited wavepacket cannot be observed, because the coherence in the period $t=[0,t_{1}]$ cannot be detected, unlike in the case of 2D correlation spectroscopy.

\FloatBarrier

\section{Conclusion}
\label{sec:conclusion}
We used the MS-QHFPE to investigate the dynamics of photoisomerization processes in a condensed phase through linear and nonlinear spectroscopy.
The linear absorption, transient absorption and two-dimensional correlation spectra were calculated for underdamped and overdamped cases.
It was shown that the two-dimensional correlation spectrum is more informative than the transient absorption spectrum, because the two-dimensional correlation spectrum can utilize the coherence among the excitation and detection periods.
For example, using the two-dimensional correlation spectrum, we can detect decay of correlation by effects of system-bath interaction as time dependence of a shape of peaks and the gradient of the excited state potential as ``coherent fringe''.

In this paper, we restricted our analysis to a system described by a single coordinate driven by a linear system-bath interaction.
Using our formulation, it is possible to investigate the effects of a non-linear system-bath interaction that causes vibrational dephasing.
Moreover, taking advantage of the computational power provided by GPGPU, it is also possible to calculate two-dimensional spectra for multi-mode anharmonic systems, for which the conical intersection plays a role \cite{chen2016fd}.
We leave such extensions to future studies to be carried out in the context of the phenomenon of molecular switching.

\begin{acknowledgments}
  Y.~T.~is thankful for financial support from a Grant-in-Aid for Scientific Research (A26248005) from the Japan Society for the Promotion of Science (JSPS).
  T.~I.~is supported by the Research Fellowships of the JSPS and a Grant-in-Aid for JSPS Fellows (16J10099).
\end{acknowledgments}

\appendix

\section{Coherent fringe}
\label{sec:coherent_fringe}
\begin{SCfigure*}
  \includegraphics[scale=0.7]{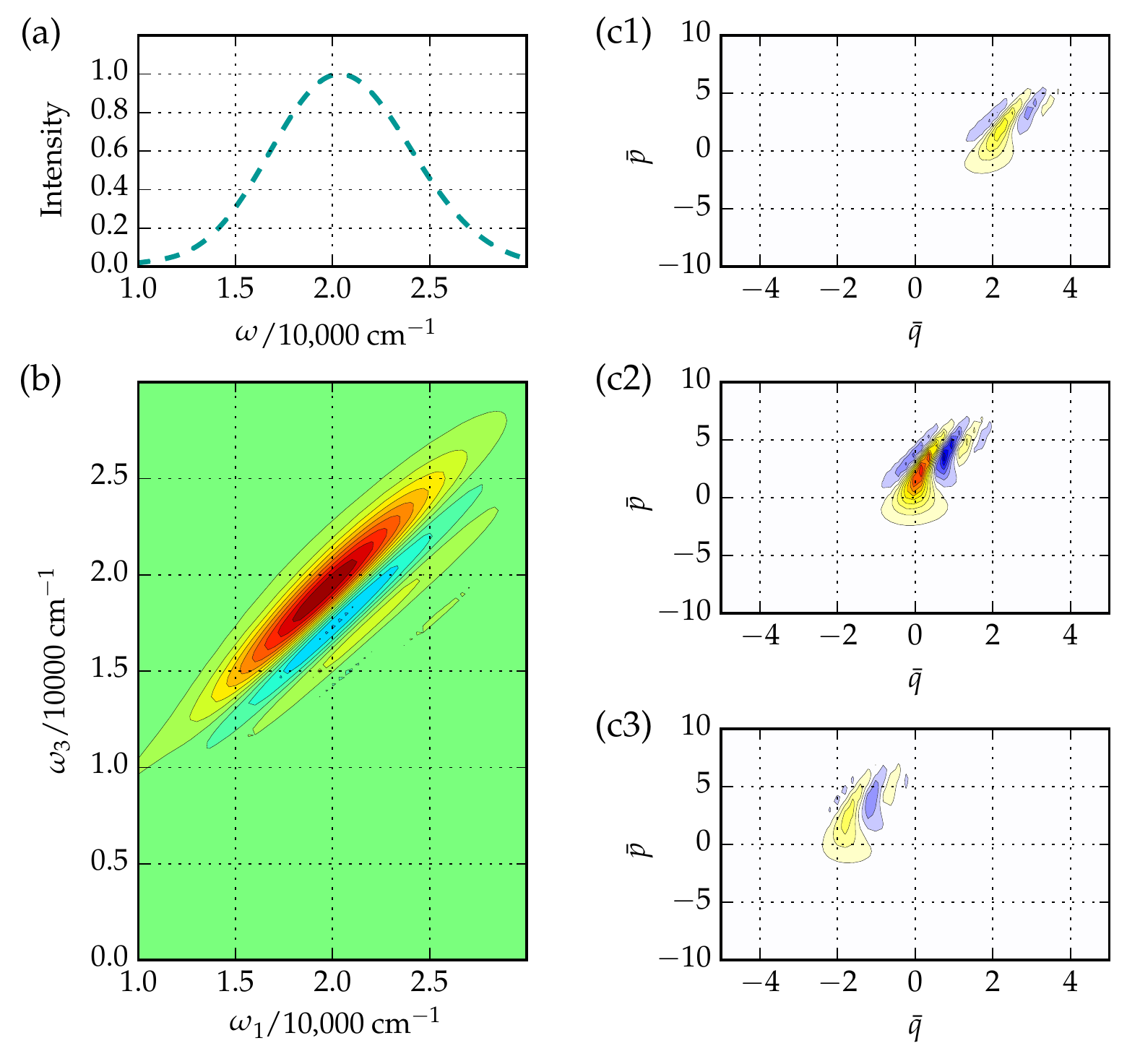}
  \caption{ (a) Linear absorption spectrum, $I^{\mathrm{abs}}(\omega )$, calculated under underdamped conditions.
    (b) The two-dimensional correlation spectrum, $I^{\mathrm{(corr)}}(\omega _{3},t_{2}=0,\omega _{1})$, calculated under underdamped conditions (same as Fig.~\ref{fig:multi_corr}(a-i)).
    The red and blue areas represent emission and absorption, respectively.
    (c) The first-order perturbed distributions of excited states, $W_{11}'(p,q,\omega _{1})$, for various different frequencies:
    (1) $\omega _{1}=15,000 \mathrm{cm}^{-1}$, (2) $\omega _{1}=20,000 \mathrm{cm}^{-1}$,(3) $\omega _{1}=25,000 \mathrm{cm}^{-1}$.
    The red and blue areas represent the positive and negative values of the partial Wigner distributions.}
  \label{fig:linear2}
\end{SCfigure*}
In order to analyze the coherent fringe appearing in Fig.~\ref{fig:multi_corr}(a-i), we introduce Wigner distributions that involve the first-order and second-order laser interactions,
\begin{widetext}
  \begin{align}
    \hat{\bm{W}}'(\omega _{1})&=\int _{0}^{\infty }\mathrm{d}t_{1}\frac{i}{\hbar }\left[\hat{\bm{\mu }}^{\rightarrow }_{W}e^{i\omega _{1}t_{1}}(\bm{\mathcal{G}}(t_{1})\hat{\bm{\mu }}_{W}^{\times }\hat{\bm{W}}_{\mathrm{eq}})-\hat{\bm{\mu }}^{\leftarrow }_{W}(\bm{\mathcal{G}}(t_{1})\hat{\bm{\mu }}^{\times }_{W}\hat{\bm{W}}_{\mathrm{eq}})e^{-i\omega _{1}t_{1}}\right]\\
    \intertext{and}
    \hat{\bm{W}}''(t_{2},\omega _{1})&=\bm{\mathcal{G}}(t_{2})\frac{i}{\hbar }\left[\hat{\bm{W}}'(\omega _{1})+\hat{\bm{W}}'(-\omega _{1})\right].
  \end{align}
\end{widetext}
With these distributions, we can express the 1D absorption and 2D correlation spectra as
\begin{align}
  &I^{\mathrm{(abs)}}(\omega _{1})=\frac{\omega }{2}\mathrm{Tr}\left\{\hat{\bm{W}}'(\omega _{1})\right\}
  \label{eq:I1W1}\\
  \intertext{and}
  &
  \begin{aligned}
    I^{\mathrm{(corr)}}(\omega _{3},t_{2},\omega _{1})&=\mathrm{Im}\int _{0}^{\infty }\mathrm{d}t_{3}e^{i\omega _{3}t_{3}}\\
    &\quad \times \mathrm{Tr}\left\{\hat{\bm{\mu }}\bm{\mathcal{G}}(t_{3})\frac{i}{\hbar }\hat{\bm{\mu }}_{W}^{\times }\hat{\bm{W}}''(t_{2},\omega _{1})\right\}.
  \end{aligned}
  \label{eq:I3W2}
\end{align}
These allows us to study the dynamics of wavepackets through spectra.
The correlation spectrum, $I^{\mathrm{(corr)}}(\omega _{3},t_{2},\omega _{1})$, can be regarded as the linear absorption spectrum of $\hat{\bm{W}}''(t_{2},\omega _{1})$.

In Fig.~\ref{fig:linear2}, 1D and 2D correlation spectra for the underdamped case are replotted from Figs.~\ref{fig:linear} and \ref{fig:multi_corr}(a-i).
A broadened absorption peak near $\omega _{1}= 20,000~\mathrm{cm}^{-1}$ is observed in the linear spectrum, while positive and negative elongated diagonal peaks centered near $(\omega _{1},\omega _{3})= (20,000~\mathrm{cm}^{-1},20,000~\mathrm{cm}^{-1})$ are observed in the 2D correlation spectrum.
The elongated peaks arise from the non-oscillatory excited wavepacket motion in the period $t=[0,t_{1}]$.
Because $t_{1}$ is finite in 2D correlation measurements, the position of the excited wave function created by the laser pulse at $t=0$ changes due to the gradient of the excited state potential when the second excitation is created by the laser pulse at $t=t_{1}$.
The difference between the locations of the first and second excited wave functions changes the phase of a density matrix element.
This is observed as a difference in momentum in Wigner space.
The small fringe-like structure of $\hat{\bm{W}}'(p,q,\omega _{1})$ depicted in Fig.~\ref{fig:linear}(c) arises from the fast motion of the excited wave function in the period $t=[0,t_{1}]$.
Because the distance between the first and second wavepackets depends on the period $t=[0,t_{1}]$, the period of the fringe-like structure varies as a function of $\omega _{1}$.
This fringe-like structure then appears as elongated positive and negative peaks in the 2D correlation spectrum, because it can be regarded as the linear absorption spectrum of $\hat{\bm{W}}''(t_{2},\omega _{1})$.
The distance between the positive and negative peaks becomes large when the motion of the first and second wave functions becomes large.
Thus, we can estimate the gradient of the potential from the sequence of these elongated peaks.

\FloatBarrier

\let\emph=\textit

\bibliography{ikeda_JCP2017}

\end{document}